\documentclass{article}
\usepackage[OT1]{fontenc} 
\usepackage{spconf,amsmath,amssymb,amsfonts,graphicx}
\usepackage{cite}
\usepackage{algorithm}
\usepackage{algorithmic}
\usepackage{enumerate}
\usepackage{mathrsfs}

\usepackage{fancyhdr}
\usepackage{color}

\usepackage{etoolbox}
\patchcmd\thebibliography
 {\labelsep}
 {\labelsep\itemsep=-1pt\relax}
 {}
 {\typeout{Couldn't patch the command}}

\title{Deep Griffin--Lim Iteration}
\name{Yoshiki Masuyama$^\dagger$, Kohei Yatabe$^\dagger$, Yuma Koizumi$^\ddag$, Yasuhiro Oikawa$^\dagger$, Noboru Harada$^\ddag$\vspace{-2pt}}
\address{
\!\!$^\dagger${\fontsize{11pt}{0pt}\selectfont Department of Intermedia Art and Science, Waseda University, Tokyo, Japan}\\
\!\!$^\ddag${\fontsize{11pt}{0pt}\selectfont NTT Media Intelligence Laboratories, Tokyo, Japan}
}
\begin{document}
\ninept
\maketitle
\begin{abstract}
\vspace{-2pt}
This paper presents a novel phase reconstruction method (only from a given amplitude spectrogram) by combining a signal-processing-based approach and a deep neural network (DNN).
To retrieve a time-domain signal from its amplitude spectrogram, the corresponding phase is required.
One of the popular phase reconstruction methods is the Griffin--Lim algorithm (GLA), which is based on the redundancy of the short-time Fourier transform.
However, GLA often involves many iterations and produces low-quality signals owing to the lack of prior knowledge of the target signal.
In order to address these issues, in this study, we propose an architecture which stacks a sub-block including two GLA-inspired fixed layers and a DNN.
The number of stacked sub-blocks is adjustable, and we can trade the performance and computational load based on requirements of applications.
The effectiveness of the proposed method is investigated by reconstructing phases from amplitude spectrograms of speeches.
\vspace{-2pt}

\end{abstract}
\vspace{-4pt}

%----------------------------------------------------
\vspace{-4pt}
\section{Introduction}
\vspace{-4pt}
%----------------------------------------------------
In recent years, phase reconstruction has gained much attention in the signal processing community \cite{Phase1, Phase2}.
Many ordinary speech processing methods defined in the time-frequency domain have considered only amplitude spectrograms and utilized the phase of the observed signal without modifying it.
Meanwhile, recent studies have proven that phase reconstruction can improve the quality of the reconstructed signal \cite{PhaseImp}, and thus several methods have been proposed for that \cite{STFTPI, Wakabayashi, Masuyama, Yatabe, MasuyamaLett}.
Phase reconstruction solely from an amplitude spectrogram has also received increasing attention along the development of the short-time Fourier transform (STFT)-based speech synthesis \cite{Takaki, Saito} which generates an amplitude spectrogram and requires phase reconstruction for generating a time-domain signal.
This paper focuses on such a situation where only an amplitude spectrogram is available for reconstructing the phase.

When only an amplitude spectrogram is available and no explicit information is given for the phase, such as in STFT-based speech synthesis, the Griffin--Lim algorithm (GLA) is one of the popular methods for phase reconstruction \cite{GLA}.
GLA promotes the consistency of a spectrogram by iterating two projections (see Section~\ref{sec: GLA}), where a spectrogram is said to be consistent when its inter-bin dependency owing to the redundancy of STFT is retained \cite{Consistency2}.
GLA is based only on the consistency and does not take any prior knowledge about the target signal into account.
Consequently, GLA often requires many iterations and results in low-quality signals.
 
For incorporating prior knowledge of target signals into phase reconstruction, deep neural networks (DNNs) have been applied recently \cite{Takahashi, Pbook, Takamichi, Oyamada}.
There exist a number of approaches to reconstruct phase using DNNs.
One approach is to treat it as a classification problem by discretizing the candidates of phase \cite{Takahashi, Pbook}, which is effectively utilized in speech separation.
Other approaches handle phase as a continuous periodic variable \cite{Takamichi} or treat complex-valued spectrogram \cite{Oyamada}.
While these DNN-based phase reconstruction methods have obtained successful results, the number of layers is determined when they are trained.
That is, their performance and computational load are fixed at the training.
It should be beneficial if one can easily trade the performance and computational load at the time of inference depending on requirements of applications.

\begin{figure}[t]
\centering
\includegraphics[width=0.92\columnwidth]{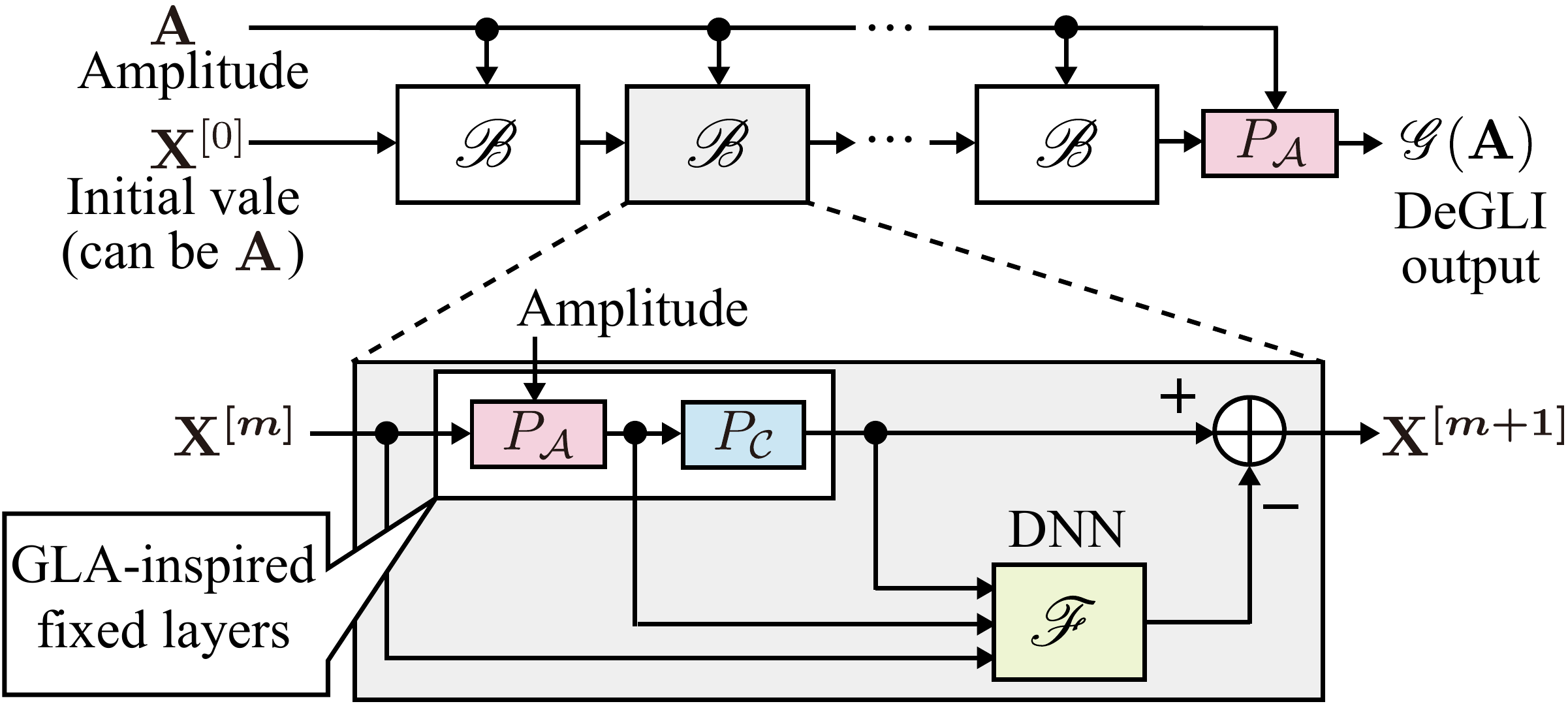}
\vspace{-8pt}
\caption{A block diagram of the proposed architecture for reconstructing phase from a given amplitude spectrogram (top), which stacks a common sub-block (bottom). The sub-block consists of two fixed GLA-inspired layers (red, blue) and a trainable DNN (green).}
\label{fig: all-architecture}
\vspace{-4pt}
\end{figure}

In this study, we propose a phase reconstruction method which incorporates a DNN into GLA.
The proposed method stacks a common sub-block motivated by the iterative procedure of GLA, which constructs a deep architecture, named \textit{deep Griffin--Lim iteration} (DeGLI),  as illustrated in Fig.~\ref{fig: all-architecture}.
In the proposed architecture, the number of total layers corresponds to the number of stacking, and its depth can be adjusted afterward based on the allowable computational load in applications.
Its training procedure is also proposed to effectively train the DNN within the sub-block.
Our main contributions are twofold: (1) proposing a deep architecture whose sub-block contains the fixed GLA-inspired layers which enable reduction of the amount of trainable parameters (Section~\ref{sec: DGLI}); and (2) proposing its training procedure which instructs the sub-block to be a denoiser, instead of requiring it to reconstruct the phase (Section~\ref{sec: Train}).
Thanks to this training procedure, the difficulty of training a DNN in phase reconstruction arisen from the periodic nature of phase is circumvented.
To evaluate the effectiveness of the proposed method, the quality of the reconstructed signal by GLA and the proposed method is compared.

%----------------------------------------------------
\vspace{-4pt}
\section{Related Works}
\vspace{-4pt}
%----------------------------------------------------

%----------------------------------------------------
\subsection{Griffin--Lim Algorithm (GLA)}
\label{sec: GLA}
\vspace{-2pt}
%----------------------------------------------------

GLA is a popular phase recovery algorithm based on the consistency of a spectrogram \cite{GLA}.
This algorithm expects to recover a complex-valued spectrogram, which is consistent and maintains the given amplitude $\mathbf{A}$, by the following alternative projection procedure:
\begin{equation}
\mathbf{X}^{[m+1]} = P_\mathcal{C} \bigl( P_\mathcal{A} (\mathbf{X}^{[m]})\bigr),
\label{eq:GLAiterDef}
\end{equation}
where $\mathbf{X}$ is a complex-valued spectrogram updated through the iteration, $P_\mathcal{S}$ is the metric projection onto a set $\mathcal{S}$, and $m$ is the iteration index.
Here, $\mathcal{C}$ is the set of consistent spectrograms, and $\mathcal{A}$ is the set of spectrograms whose amplitude is the same as the given one.
The metric projections onto these sets $\mathcal{C}$ and $\mathcal{A}$ are given by,
\begin{align}
P_\mathcal{C}(\mathbf{X}) &= \mathcal{G} \mathcal{G}^{\dagger} \mathbf{X}, \label{eq:PC}\\[3pt]
P_\mathcal{A}(\mathbf{X}) &= \mathbf{A} \odot \mathbf{X} \oslash |\mathbf{X}|,
\end{align}
where $\mathcal{G}$ represents STFT, $\mathcal{G}^\dagger$ is the pseudo inverse of STFT (iSTFT), $\odot$ and $\oslash$ are element-wise multiplication and division, respectively, and division by zero is replaced by zero.
GLA is obtained as an algorithm for the following optimization problem \cite{Consistency2}:
\begin{equation}
\min_\mathbf{X}  \,\,\, \left\|  \mathbf{X} - P_\mathcal{C}(\mathbf{X}) \right\|_\text{Fro}^2 \,\,\,\,
\text{s.t.} \,\,\,\,\, \mathbf{X} \in \mathcal{A},
\label{eq: GLAOpt}
\end{equation}
where $\| \cdot \|_\text{Fro}$ is the Frobenius norm.
This equation minimizes the energy of the inconsistent components under the constraint on amplitude which must be equal to the given one.
Although GLA has been widely utilized because of its simplicity, GLA often involves many iterations until it converges to a certain spectrogram and results in low reconstruction quality.
This is because the cost function in Eq.~\eqref{eq: GLAOpt} only requires the consistency, and the characteristics of the target signal are not taken into account. 
Introducing prior knowledge of the target signal into the algorithm can improve the quality of reconstructed signals as discussed in \cite{CAWF, Yatabe}.

%----------------------------------------------------
\vspace{-2pt}
\subsection{DNN-based phase reconstruction with fixed STFT layers}
\vspace{-2pt}
%----------------------------------------------------

Recently, DNNs including fixed STFT (and iSTFT) layers were considered for treating phase information within the networks.
A generative adversarial network (GAN)-based approach to reconstruct a complex-valued spectrogram solely from a given amplitude spectrogram was presented in \cite{Oyamada}.
The output of the generator (a complex-valued spectrogram) is converted back to the time domain by  iSTFT layer and inputted to the discriminator, where this iSTFT layer is essential for its training as discussed in \cite{Oyamada}.
As another example, a DNN for speech separation \cite{MERL1} employed the multiple input spectrogram inverse (MISI) layer which consists of the pair of STFT and iSTFT as in GLA.
The MISI layer is applied to the output of the DNN for speech separation to improve its performance by considering the effect of the phase reconstruction together with the separation.
In addition, in \cite{MERL2}, the time-frequency representation was also trained with the DNN for speech separation.
The success of these DNNs indicates that considering STFT (and iSTFT) together with a DNN is important for treating phase.

The common strategy for these DNNs is that fixed STFT-related layers are placed after a rich DNN.
Their loss functions are evaluated after going through such STFT-related layers, and their effect is propagated for updating the parameters of DNNs.
Based on this observation, loss functions tied with STFT (and iSTFT) seem important in phase reconstruction because such loss functions are related to the concept of the consistency.
At the same time, fixed STFT-related layers have several benefits for training.
Since they do not contain trainable parameters, adding STFT-related layers does not increase the number of trainable parameters while they capture the structure of complex-valued spectrograms efficiently.
Therefore, use of the STFT-related layers within DNNs may be recommended for treating phase information.
However, there are little research on such DNN containing STFT within the network.

%----------------------------------------------------
\vspace{-4pt}
\section{Proposed deep architecture}
\vspace{-4pt}
%----------------------------------------------------

Based on the above discussions, we propose an architecture for phase reconstruction, named deep Griffin--Lim iteration (DeGLI), which is a unification of GLA and a DNN.
As illustrated in Fig.~\ref{fig:  all-architecture}, the proposed architecture consists of a common sub-block, and it is stacked to form the whole deep architecture based on the iterative procedure of GLA.
The architecture of DeGLI is introduced in Section~\ref{sec: DGLI}, while its training procedure is described in Section~\ref{sec: Train}.

%----------------------------------------------------
\vspace{-4pt}
\subsection{Deep Griffin--Lim Iteration (DeGLI)}
\label{sec: DGLI}
\vspace{-2pt}
%----------------------------------------------------

One interesting trend of research in deep learning is to interpret an optimization algorithm as a recurrent neural network (RNN) and construct a DNN architecture following that \cite{LISTA, LAMP, ADMMNet}.
The DNN introduced in the previous section \cite{MERL1, MERL2} was also obtained by a similar approach called deep unfolding \cite{UnF1, UnF2}.
In this context, the iterative procedure of GLA in Eq.~\eqref{eq:GLAiterDef} is interpreted as an RNN which stacks the fixed linear layer $P_\mathcal{C}$ and target-dependent nonlinear layer $P_\mathcal{A}$.
By looking close at Eq.~\eqref{eq:GLAiterDef}, it can be seen that the complex-valued spectrogram at $m$th iteration $\mathbf{X}^{[m]}$ is inputted into the nonlinear layer $P_\mathcal{A}$, and then its output passes through the fixed linear layer $P_\mathcal{C}$ consisting of STFT $\mathcal{G}$ and iSTFT $\mathcal{G}^\dagger$ as in Eq.~\eqref{eq:PC}.
That is, GLA is a parameter-fixed RNN consisting of STFT and iSTFT layers within the network.
Inspired from the above observations, the proposed deep architecture for phase reconstruction, or DeGLI, is defined through a sub-block based on GLA.

Let us consider the intermediate representation of GLA,
\begin{align}
    \mathbf{Y}^{[m]} &= P_\mathcal{A}(\mathbf{X}^{[m]}),\\[1pt]
    \mathbf{Z}^{[m]} &= P_\mathcal{C}(\mathbf{Y}^{[m]}),
\end{align}
where the combination of these equations recovers Eq.~\eqref{eq:GLAiterDef}.
Since $\mathbf{Y}^{[m]}$ is the amplitude-replaced version of $\mathbf{X}^{[m]}$, their difference indicates the amount of mismatch between the amplitude of current spectrogram $|\mathbf{X}^{[m]}|$ and the desired amplitude $\mathbf{A}$.
Similarly, since $\mathbf{Z}^{[m]}$ is the closest consistent spectrogram to $\mathbf{Y}^{[m]}$ (in the Euclidean sense), the difference between them indicates the amount of inconsistent components \cite{Consistency2}.
Such differences should be quite informative for phase reconstruction because the aim of GLA is to reduce them as much as possible.
However, such intermediate information is not considered in the original GLA in Eq.~\eqref{eq:GLAiterDef}.

To effectively use those intermediate information in a learning scheme, we propose DeGLI as the following architecture:
\begin{align} 
\mathbf{X}^{[m+1]} & = \mathscr{B} (\mathbf{X}^{[m]}), \label{eq:DeGLIdef}\\
& = \mathbf{Z}^{[m]} - \mathscr{F} (\mathbf{X}^{[m]}, \mathbf{Y}^{[m]}, \mathbf{Z}^{[m]}),
\label{eq:DeGLIdef2}
\end{align}
where $\mathscr{B}$ is the proposed DeGLI-block inspired by GLA as in Fig.~\ref{fig:  all-architecture}, and $\mathscr{F}$ is a DNN.
The whole architecture can also be viewed as an RNN or a feed-forward network in which the weights are shared.
By stacking $M$ DeGLI-blocks (which is equivalent to iterate Eq.~\eqref{eq:DeGLIdef} $M$ times), the whole DeGLI architecture becomes $M$-times deeper without increasing the number of trainable parameters. 
That is, the total depth of the DeGLI architecture can be adjusted afterward, which enables one to easily trade its performance and computational load for adapting the allowable computational time of various applications.
Note that, as a specific case, DeGLI reduces to the ordinary GLA when $\mathscr{F} (\mathbf{X}^{[m]}, \mathbf{Y}^{[m]}, \mathbf{Z}^{[m]}) = \mathbf{O}$, where $\mathbf{O}$ is the zero matrix.
A variant of GLA in \cite{FGLA} can also be obtained by setting $\mathscr{F} (\mathbf{X}^{[m]}, \mathbf{Y}^{[m]}, \mathbf{Z}^{[m]}) = \gamma ( \mathbf{X}^{[m]} -  \mathbf{Z}^{[m]}) \;$ $(0<\gamma<1)$, which indicates that DeGLI is a general architecture including several GLA-type algorithms as spacial cases.

One of the key points of the DeGLI architecture is that $\mathbf{Z}^{[m]}$ $(=\!P_\mathcal{C} ( P_\mathcal{A} (\mathbf{X}^{[m]})))$ is the output of GLA-inspired layers at $m$th iteration, and the proposed DeGLI-block $\mathscr{B}$ is defined as the subtraction of the DNN output $\mathscr{F} (\mathbf{X}^{[m]}, \mathbf{Y}^{[m]}, \mathbf{Z}^{[m]})$ from the output of GLA-inspired layers $\mathbf{Z}^{[m]}$.
Defining the DeGLI-block in this way is based on two reasons: (1) differences between each intermediate representation indicates the \textit{residual} to the desired ones as discussed in the above paragraph; and (2) it is known that treating a residual is easier than directly estimating the target according to the literature on the residual learning \cite{ResNet, DnCNN}.

\begin{figure}[t]
	\centering
	\includegraphics[width=0.92\columnwidth]{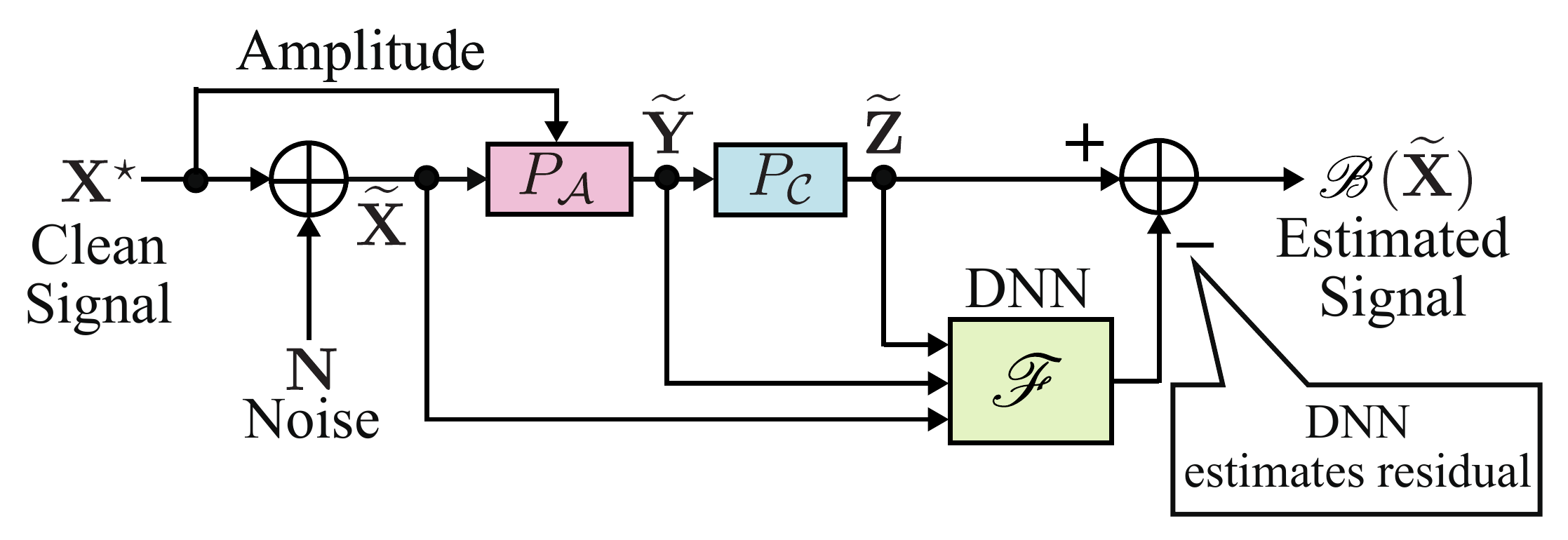}
	\vspace{-12pt}
	\caption{The block diagram for training the sub-block.}
\label{fig: Train}
\vspace{-4pt}
\end{figure}

%----------------------------------------------------
\vspace{-4pt}
\subsection{Training procedure for DeGLI-block}
\label{sec: Train}
\vspace{-2pt}
%----------------------------------------------------
Since the proposed DeGLI architecture can be interpreted as a large DNN, a simple strategy for training the DeGLI-block is directly minimizing the loss of phase reconstruction measured by the output:
\begin{equation}
\vspace{-2pt}
\min_{\theta} \,\, D \bigl( \mathscr{G}_\theta (\mathbf{A}), P_\mathcal{C} ( \mathscr{G}_\theta (\mathbf{A}) ) \bigr),
\vspace{-2pt}
\end{equation}
where $\mathscr{G}(\cdot) = P_\mathcal{A}(\mathscr{B}(\cdots\mathscr{B}(\mathscr{B}(\cdot))))$ represents the whole DeGLI architecture, $\theta$ represents all trainable parameters in $\mathscr{G}$ (i.e., the parameters in $\mathscr{F}$), $D(\cdot, \cdot)$ is a measure of mismatch such as a norm of difference, and the minimization is considered for all $\mathbf{A}$.
This problem is related to the optimization problem for GLA in Eq.~\eqref{eq: GLAOpt} when $D$ is the squared Frobenius norm of the difference of the variables (note that, since $P_\mathcal{A}$ is applied at the last of $\mathscr{G}$, the constraint in Eq.~\eqref{eq: GLAOpt} is always satisfied).
Although the above training strategy is straightforward, the number of the blocks should be defined in advance for applying it.
In addition, it did not work well in our preliminary experiments.

In order to tackle this issue, we train the DeGLI-block $\mathscr{B}$ to be a denoiser by the training procedure illustrated in Fig.~\ref{fig: Train}.
Let $\mathbf{X}^\star$ be a complex-valued spectrogram of a target signal, and $\widetilde{\mathbf{X}} = \mathbf{X}^\star\! + \mathbf{N}$ be its noisy counterpart degraded by complex-valued noise $\mathbf{N}$.
Then, the DeGLI-block $\mathscr{B}$ is trained so that $\mathscr{B}(\widetilde{\mathbf{X}})\approx\mathbf{X}^\star$, i.e.,
\begin{equation}
\vspace{-2pt}
    \mathscr{B}(\widetilde{\mathbf{X}}) = \widetilde{\mathbf{Z}}-\mathscr{F} (\widetilde{\mathbf{X}}, \widetilde{\mathbf{Y}}, \widetilde{\mathbf{Z}}) \approx\mathbf{X}^\star,
\vspace{-2pt}
\end{equation}
based on the definition in Eq.~\eqref{eq:DeGLIdef2}.
Since $\widetilde{\mathbf{Z}}$ is obtained only from the fixed layers $P_\mathcal{A}$ and $P_\mathcal{C}$, the optimization problem for training the DNN $\mathscr{F}$ is given by
\begin{equation}
\vspace{-2pt}
\min_{\theta} \,\, D \bigl( \widetilde{\mathbf{Z}} - \mathbf{X}^\star, \mathscr{F}_\theta (\widetilde{\mathbf{X}}, \widetilde{\mathbf{Y}}, \widetilde{\mathbf{Z}}) \bigr).
\label{eq: Loss}
\vspace{-2pt}
\end{equation}
In such denoising, the DNN $\mathscr{F}$ estimates the residual components $\widetilde{\mathbf{Z}} - \mathbf{X}^\star$ which should not be contained in the GLA output $\widetilde{\mathbf{Z}} = P_\mathcal{C} ( P_\mathcal{A} (\widetilde{\mathbf{X}}))$.
To be specific,  the DNN takes the mismatch to the consistency and amplitude into account by inputting $\widetilde{\mathbf{Y}}$ and $\widetilde{\mathbf{Z}}$, and it implicitly eliminates the latent target signal (such as clean speech) through hidden layers in $\mathscr{F}$.
This training strategy is closely related to the residual learning strategy.
It has been shown that a denoising sub-block with the residual learning strategy is robust to the type and level of noise, and it can be applied to a variety of tasks as discussed in \cite{DnCNN, ResApp1}.
The idea of applying a denoising DNN for general tasks can also be found in \cite{PPP1, RED1}.

Note that, after passing through the fixed nonlinear layer $P_\mathcal{A}$, the amplitude of the complex-valued spectrogram is always replaced by the desired one.
That is, the difference between $\widetilde{\mathbf{Y}}$ and the target $\mathbf{X}^\star$ is only phase, and thus denoising of $\widetilde{\mathbf{Y}}$ $(=\!P_\mathcal{A} (\widetilde{\mathbf{X}}))$ corresponds to phase reconstruction.
It can be expected that the denoising sub-block including GLA-inspired layers also works well in phase reconstruction.
In any case, the trained DNN $\mathscr{F}$ (and thus $\mathscr{B}$) only affects the phase of the final output because the amplitude is always set to the given one by $P_\mathcal{A}$ after the last DeGLI block.

\begin{figure}[t]
	\centering
	\includegraphics[width=0.92\columnwidth]{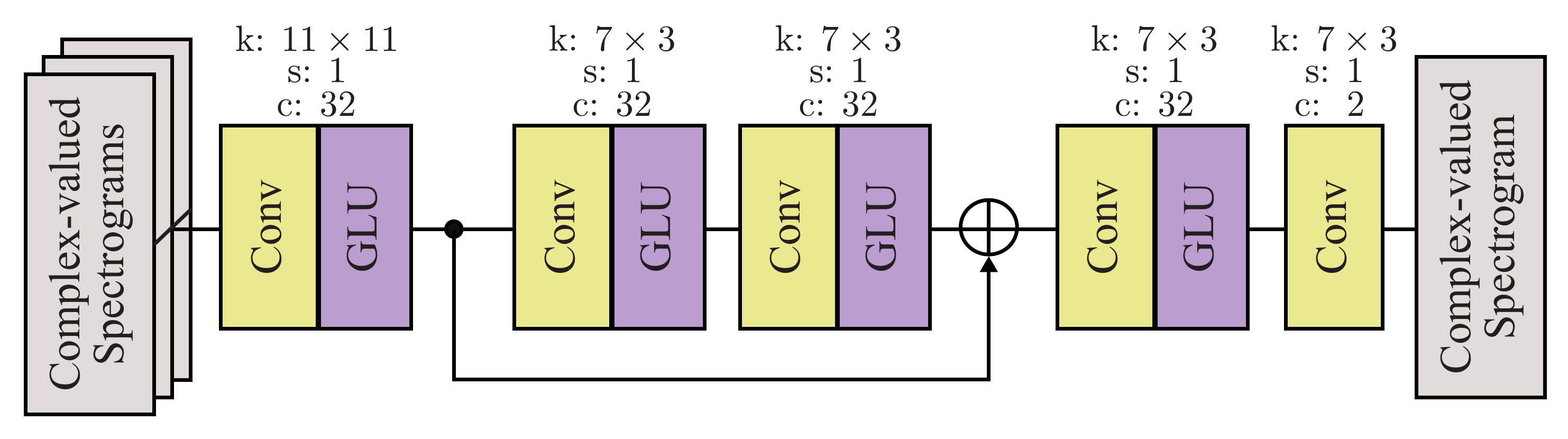}
	\vspace{-12pt}
	\caption{The illustration of the DNN used in the experiment.
It maps real and imaginary parts of three complex-valued spectrograms ($\mathbf{X}$, $\mathbf{Y}$, and $\mathbf{Z}$) to those of the residual.
Here, ``Conv'' indicates a convolutional layer with the zero padding for keeping the input size, where $\mathrm{k}$, $\mathrm{s}$, and $\mathrm{c}$ are the kernel size, stride size, and the number of channels, respectively.
``GLU'' represents the gated linear unit.}
\label{fig: Net}
\vspace{-4pt}
\end{figure}

%----------------------------------------------------
\vspace{-4pt}
\section{Experiment}
\label{sec: Experiments}
\vspace{-4pt}
%----------------------------------------------------

In order to validate the effectiveness of DeGLI, the quality of reconstructed speeches was evaluated by objective measures.
The proposed method was compared with GLA as a baseline method.

%----------------------------------------------------
\vspace{-4pt}
\subsection{Experimental settings}
\vspace{-2pt}
%----------------------------------------------------

A DNN $\mathscr{F}$ used in the DeGLI-block $\mathscr{B}$ for the experiment is illustrated in Fig.~\ref{fig: Net}.
The $2$D Convolutional layers (Conv) and the gated linear units (GLU) \cite{GCNN} are stacked with the skip connections.
In the Conv layers, the complex-valued spectrograms are treated as images, where the real and imaginary parts are concatenated along the channel direction.
Note that the input of the DNN is three complex-valued spectrograms as in Figs.~\ref{fig: all-architecture} and \ref{fig: Train}, which results in six channels as each of the three consists of the real and imaginary parts.

As the training dataset for denoising, the Wall Street Journal (WSJ-$0$) corpus recorded at the sampling rate of $16$ kHz was utilized.
$14\, 250$ speech files were randomly selected from the database to form a training set, and the rest of the data was used as a validation set.
During the mini-batch training, the utterances were divided into about $2$-second-long segments ($32\,768$ samples), and the Adam optimizer was utilized as the optimization solver.
The network was trained for $50$ epochs with a learning rate control, where the learning rate was decayed by multiplying $10^{-0.5}$ if the loss function on the validation set did not decrease for $2$ consecutive epochs, and the initial learning rate was set to $10^{-3}$.
As the noise utilized for training in the time-frequency domain (described in Section~\ref{sec: Train}), the complex Gaussian noise was added so that the signal-to-noise ratio was randomly selected from $-6$ to $0$ dB, and the measure of mismatch $D$ as the loss function in Eq.~\eqref{eq: Loss} was set to the $\ell_1$-norm of difference.
STFT was implemented with the Hann window, whose duration was $64$ ms, with $32$ ms shifting.
As the test dataset, randomly selected $500$ utterances from the TIMIT dataset were utilized for obtaining amplitude spectrograms for phase reconstruction, where the initial phases were set to zero in the time-frequency domain (i.e., the amplitude spectrogram was directly inputted as the initial value).

%----------------------------------------------------
\vspace{-4pt}
\subsection{Experimental results}
\vspace{-2pt}
%----------------------------------------------------

\begin{figure}[t]
	\centering
	\includegraphics[width=0.92\columnwidth]{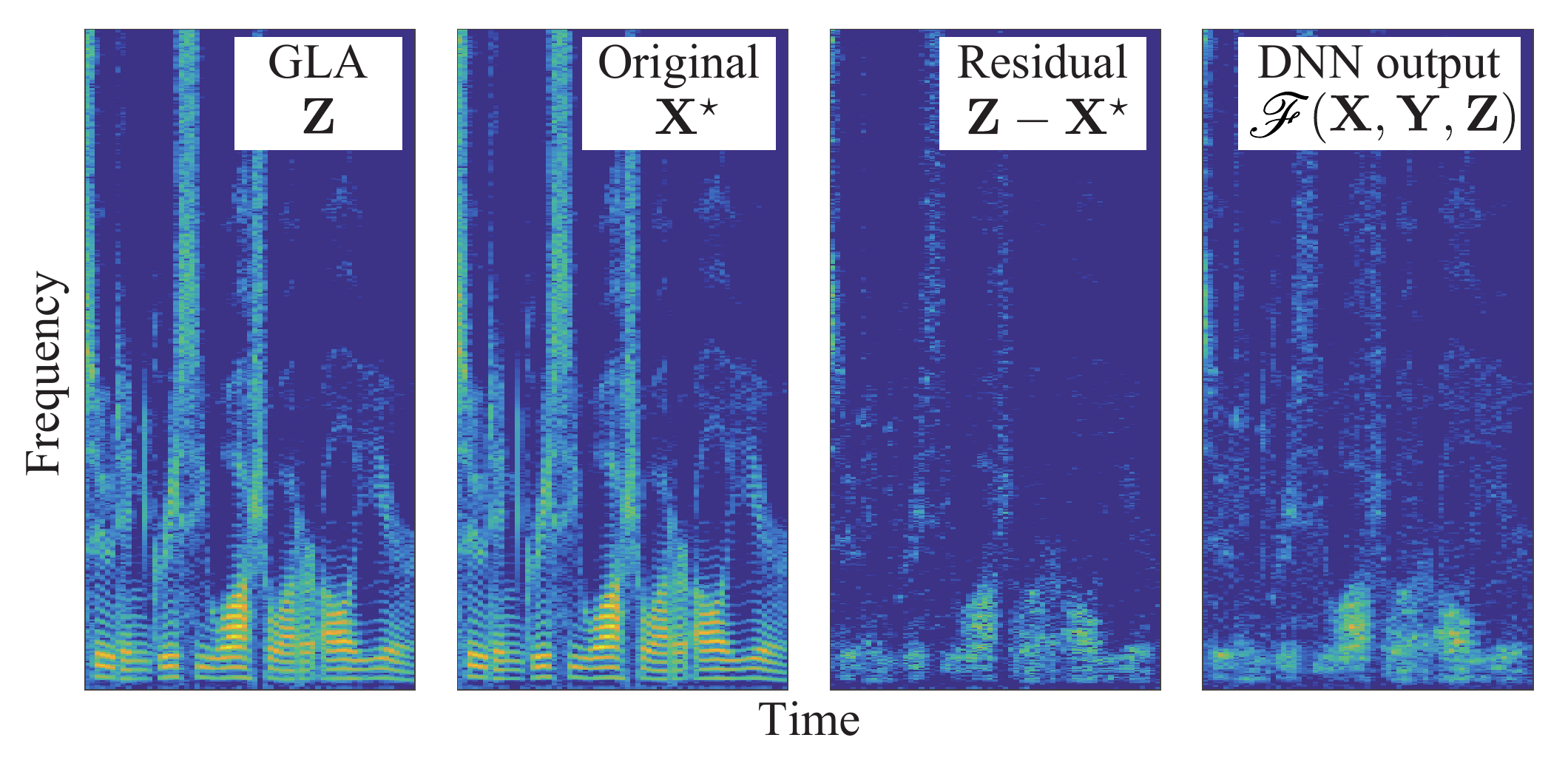}
	\vspace{-8pt}
	\caption{An example of the spectrograms within the proposed DeGLI-block. ``GLA'' represents the output of the GLA-inspired layers $\mathbf{Z} = P_\mathcal{C} (P_\mathcal{A} (\mathbf{X}))$, and ``Original'' is the clean speech signal $\mathbf{X}^\star$ to be recovered. The difference between them $\mathbf{Z}-\mathbf{X}^\star$ is shown as ``Residual'', while its estimation $\mathscr{F}(\mathbf{X},\mathbf{Y},\mathbf{Z})$ is denoted by ``DNN output''. The DNN was able to accurately estimate the Residual.}
\label{fig: Spectrogram}
\vspace{-4pt}
\end{figure}

An example of the results of the residual learning is shown in Fig.~\ref{fig: Spectrogram} for illustrating how the DNN in the proposed DeGLI-block works.
As shown in the figure, the DNN appropriately estimated the residual which is the difference between the output of the current GLA-inspired layers and the target spectrogram.
Such estimated residual is subtracted so that the difference to the ideal spectrogram is reduced.
We expect that the estimation by the DNN is reasonably accurate to improve the output of the DeGLI-block.

\begin{figure}[t]
	\centering
	\includegraphics[width=0.99\columnwidth]{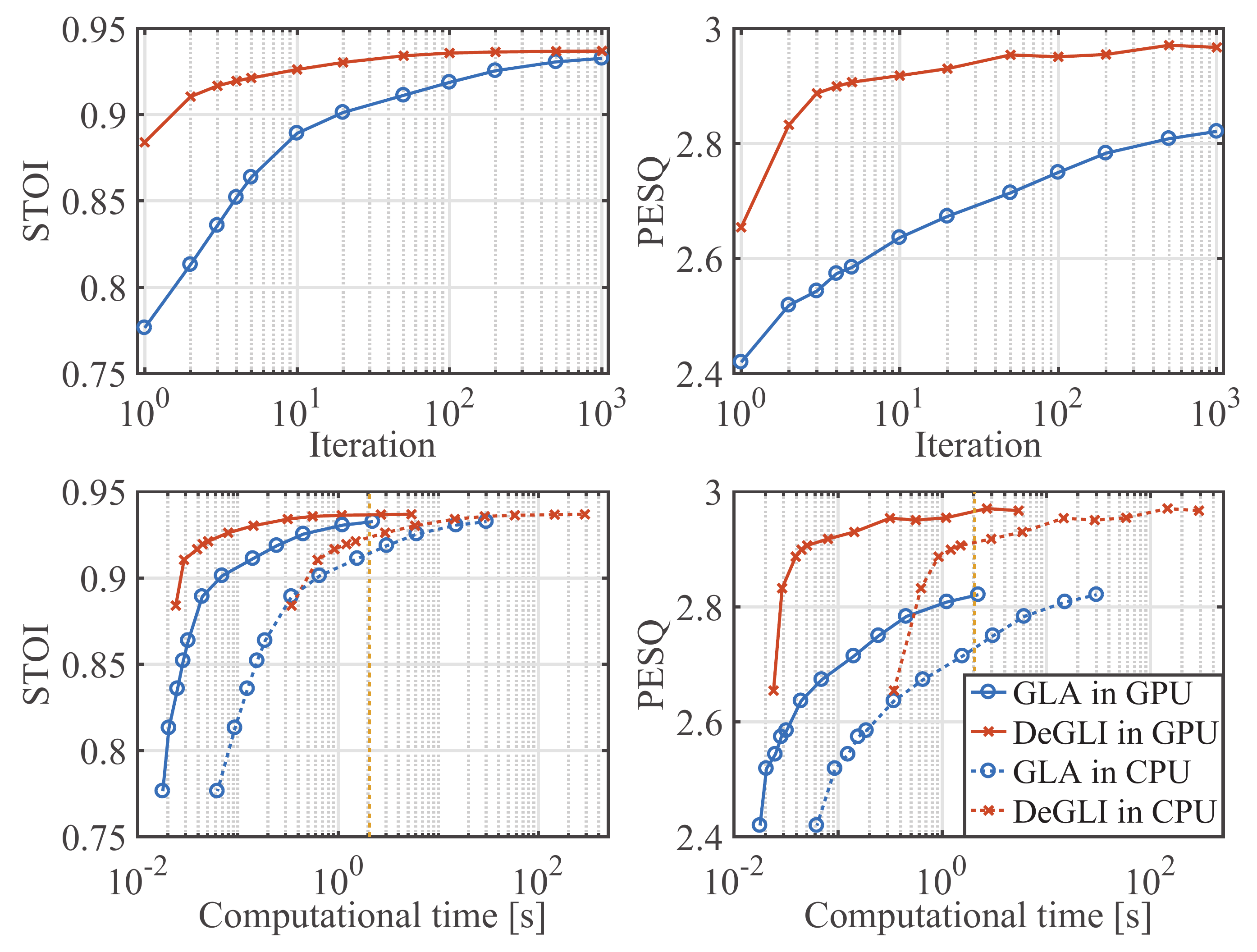}
	\vspace{-8pt}
	\caption{Average scores of STOI and PESQ per iteration (top) and per computational time (bottom) for GLA (blue, circles) and the proposed method (red, cross marks). The yellow dashed line indicates that the real time factor is $1$. For measuring the computational time, both methods were implemented by using CPU and GPU.}
\label{fig: Objective}
\vspace{-2pt}
\end{figure}

The performance of phase reconstruction was evaluated by STOI \cite{STOI} and PESQ \cite{PESQ}.
The score per iteration averaged among the test set is shown in the upper row of Fig.~\ref{fig: Objective}.
Both STOI and PESQ of the proposed method were always higher than those of GLA at each iteration, and it improved the performance as the number of iteration increased.
Since the iteration corresponds to the depth of the whole architecture of DeGLI, this result indicates that one can iterate the DeGLI-block until the quality of the reconstructed signal become satisfactory.
Namely, one can eliminate unnecessary computation, or decide the depth based on the available computational resource at that time.
We stress that this unique feature of the proposed method cannot be achieved by a single rich DNN directly mapping an inputted amplitude spectrogram into the final reconstructed signal.

\begin{figure}[t]
	\centering
	\includegraphics[width=0.99\columnwidth]{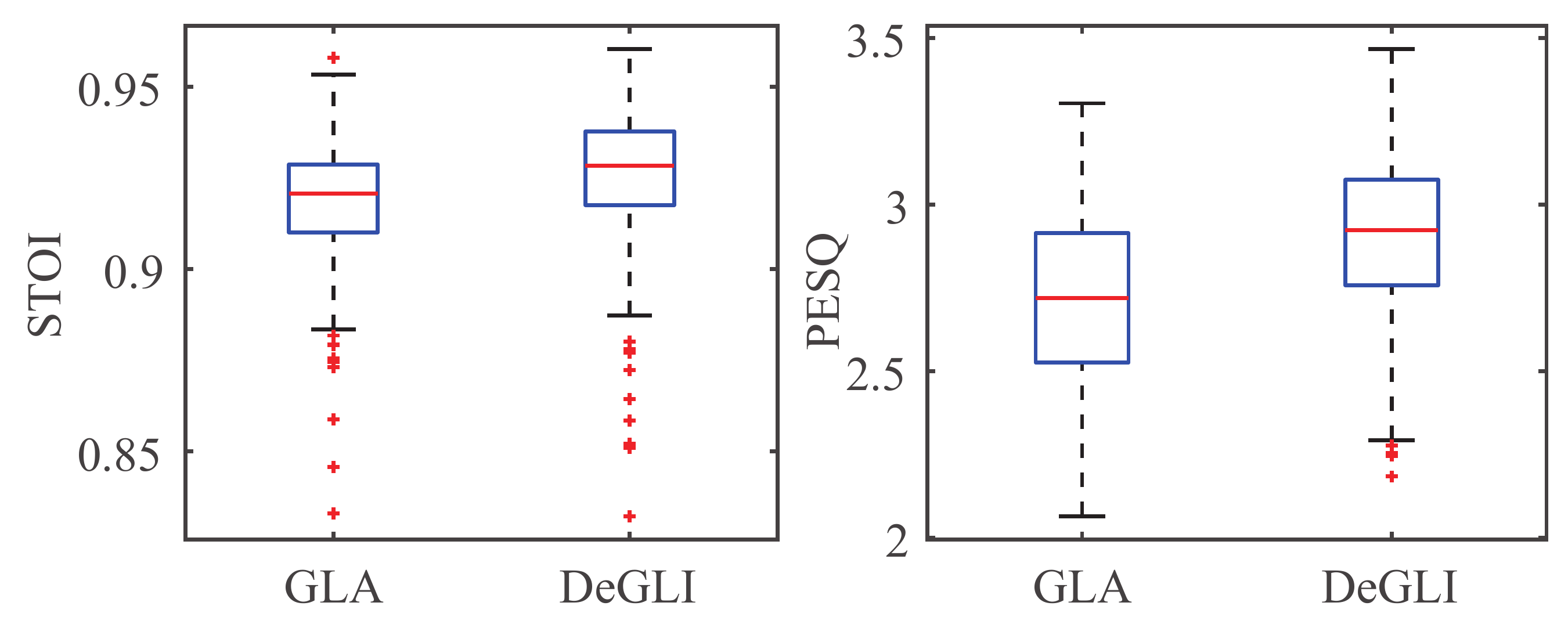}
	\vspace{-8pt}
	\caption{Box plots of the scores of STOI and PESQ, where the results of GLA were evaluated at the $100$th iteration, while those of DeGLI were obtained from the output of the $10$th stacks. The red lines are the median, and the boxes indicates the first and third quartiles.}
\label{fig: Box}
\vspace{-2pt}
\end{figure}

Since the computational time per iteration is different between GLA and the proposed DeGLI-block, the performance was also investigated in terms of computational time for fair comparison.
In this experiment, ``Intel Core i$9$-$7980$XE ($2.60$ GHz)'' and ``NVIDIA GeForce GTX $1080$ Ti'' were employed for the CPU and GPU, respectively.
For both methods, STFT and iSTFT were implemented by TensorFlow. 
The scores per computational time are illustrated in the bottom row of Fig.~\ref{fig: Objective}.
Since the computational time per iteration of the proposed method was about $2.4$ and $9.6$ times slower than GLA by using GPU and CPU, respectively, the difference of the scores between the methods is closer than in the top row.
Nevertheless, the proposed method notably outperformed GLA especially for PESQ.
To see the scores at some specific iterations, box plots of the scores are also shown in Fig.~\ref{fig: Box}.
The results were obtained from the $100$th iteration for GLA and the $10$th iteration for the proposed method because the computational times of these methods are roughly the same at those iteration numbers.
It can be seen that the tendencies of the scores are the same as the averaged values in Fig.~\ref{fig: Objective}, and the effectiveness of the proposed DeGLI architecture was confirmed by a paired one-side $t$-test ($p\!<\!0.01$).

In summary, it was confirmed that the proposed DeGLI architecture can be trained so that utilizing the common block for every iteration improves the performance, which should be because of training as a denoiser and the residual learning strategy.
Note that the trainable DNN used in this experiment was merely an example, and it must be possible to improve the performance by considering a DNN more suitable for phase reconstruction.

%----------------------------------------------------
\vspace{-4pt}
\section{Conclusion}
\vspace{-4pt}
%----------------------------------------------------

In this study, we proposed a deep architecture, named DeGLI, which combines a DNN with the iterative procedure of GLA.
The key idea was to stack the same sub-block, so that the depth of whole architecture can be adjusted without increasing the number of trainable parameters.
This feature enables one to trade the quality of the reconstructed signal and computational load depending on applications.
The residual learning strategy was applied to train the sub-block as a denoiser, where the DNN removes the undesired components introduced by GLA.
Experimental results confirmed that a denoising sub-block is applicable to phase reconstruction, which indicates that the task of training can be different from the phase reconstruction which is not an easy task for a DNN owing to the periodic nature of phase.
Investigation of a DNN suitable for the proposed DeGLI remains as a future work.

%In this study, we proposed a deep architecture, named DeGLI, which combines a DNN with the iterative procedure of GLA.
%The key idea was to stack the same sub-block (DeGLI-block), so that the depth of the proposed architecture can be adjusted without increasing the number of trainable parameters.
%This feature enables one to trade the quality of the reconstructed signal and computational load depending on applications.
%The residual learning strategy was applied to train the sub-block as a denoiser, where the DNN removes the undesired components introduced by GLA.
%Experimental results confirmed that a denoising sub-block is applicable to phase reconstruction, which should be an important finding because it indicates that the task of training can be different from the phase reconstruction which is not an easy task for a DNN owing to the periodic nature of the phase.
%Investigation of a DNN suitable for the proposed DeGLI remains as a future work.

% References should be produced using the bibtex program from suitable
% BiBTeX files (here: strings, refs, manuals). The IEEEbib.bst bibliography
% style file from IEEE produces unsorted bibliography list.
% -------------------------------------------------------------------------
\bibliographystyle{IEEEbib}

\end{document}